\newcommand\ignore[1]{}

\documentclass[11pt, oneside]{article}   	% use "amsart" instead of "article" for AMSLaTeX format
\usepackage{geometry}                		% See geometry.pdf to learn the layout options. There are lots.
\usepackage[parfill]{parskip}    		% Activate to begin paragraphs with an empty line rather than an indent
\usepackage{graphicx}				% Use pdf, png, jpg, or eps� with pdflatex; use eps in DVI mode
\usepackage{color}								% TeX will automatically convert eps --> pdf in pdflatex		
\usepackage[left]{lineno}
\usepackage{amssymb}
\usepackage{amsmath}
\usepackage{amssymb}
\usepackage{graphicx}
\usepackage{hyperref}
\usepackage{subfigure}
\usepackage[subfigure]{tocloft}
\usepackage{float}
\usepackage{fullpage}
\usepackage{braket}
\usepackage{url}
\usepackage{cite}
\usepackage{upgreek}
\usepackage{outlines}
\usepackage{braket}
\usepackage{appendix}
\usepackage{xspace}
\usepackage{tikz}
\usepackage{placeins}
\usepackage{multirow}
\usetikzlibrary{matrix,arrows,decorations.pathmorphing}
\usetikzlibrary{patterns}
\usetikzlibrary{trees}
\usetikzlibrary{decorations.pathmorphing}
\usetikzlibrary{decorations.markings}
\usetikzlibrary{positioning,arrows}
\usetikzlibrary{decorations.pathreplacing}
\usetikzlibrary{decorations.pathmorphing}
\usetikzlibrary{decorations.markings}

\newcommand\be{\begin{equation}}
\newcommand\ee{\end{equation}}
\newcommand\bea{\begin{eqnarray}}
\newcommand\eea{\end{eqnarray}}

\newcommand{\bdm}{\begin{displaymath}}
\newcommand{\edm}{\end{displaymath}}

\numberwithin{equation}{section}
\numberwithin{figure}{section}

\renewcommand{\epsilon}{\varepsilon}

\usepackage{authblk}

\renewcommand{\phi}{\varphi}

\graphicspath{{fig/,rootFigs/}}

\title{The Pomeron  --- A Bootstrap Story} 

\author[*]{Chung-I Tan}
\affil[*]{Brown University, Providence, RI 02912}

\begin{document}

\maketitle

\begin{abstract}
In a contribution to the volume {\it A Passion for Physics},  a collection of essays in honor of Geoffrey Chew's sixtieth birthday, I wrote, together with A.~Capella, Uday Sukhatme, and Tran Thanh Van {\it The Pomeron Story.}  This is a follow-up to that contribution.  This sequel~\footnote{This is a contribution to the book ``Geoff Chew: Architect of the Bootstrap'', to be published by World Sceintific.} also serves as an opportunity to acknowledge my gratitude to Geoff as a PhD student under his tutelage.
\end{abstract}
%\newpage
%\tableofcontents
%\newpage
\setlength{\parskip}{.1in}
\maketitle

\pagenumbering{gobble}
%\pagebreak
\pagenumbering{arabic}

\section{My Years at Berkeley}
In our 1984 essay, {\it The Pomeron Story}, we wrote ``In their seminal paper~\cite{Chew:1962eu} Chew and Frautschi asserted that all hadrons are composite; they lie on Regge trajectories.  $\ldots\ldots$"   To account for the near constancy of hadronic total cross sections, it was necessary for them to postulate the existence of a Regge trajectory with vacuum quantum numbers and zero-energy intercept at $j=1$,  even though no physical particles were known at the time to lie on the trajectory. This Regge singularity, which was also proposed independently by Gribov, is known as the Pomeron. 

 \fontdimen2\font=2.79pt
 Our understanding of the Pomeron has evolved since {\it The Pomeron Story} was written. The fundamental theory for the strong interactions, Quantum Chromodynamics (QCD),  has been established. However, much of its tests involved perturbative aspects of QCD. 
 Great progress has also been made via lattice studies. Yet,  near-forward high energy scattering %in the forward limit 
 remains  a challenge.  What is the Pomeron, really?
\fontdimen2\font=3.5pt 

As a student at Berkeley in the late sixties, one faced constant turmoil on social issues. We were all caught up with the Free Speech Movement, \mbox{subscribing to} \mbox{{\it Ramparts}}, etc. As a student of Geoff, we also found relief in moving daily to the Rad Lab on the hill.   Geoff had a large group of graduate students, sharing a common interest in various aspects of strong interactions, ranging from purely phenomenological to foundational aspects of S-matrix theory. This group included Ling-Lie Chau, Jerry Finkelstein, Jiunn-Ming Wang,  Farzam Arbab, Jan Dash, Huan Lee, Richard Brower, Michael Misheloff, Carleton DeTar, Dennis Sivers, Don Tow, and others.\footnote{Students just finished included David Gross, John Schwarz, Shu-Yuan Chu, John Stack,  etc. Other contemporary students include Dick Haymaker,  Lay-Nam Chang, and others.}  Together with other students, e.g., Dick Slansky and Joe Weis, and also a large  group of post-docs,  a common theme shared by most  was the question of consistency and uniqueness of the S-matrix, as exemplified by the  bootstrap program\cite{Chew:1963zza,DHS}. Can all Regge parameters~\cite{Mandelstam:1968zza} be fixed within an analytic S-matrix? One could not avoid discussing the Pomeron and Bootstrap nearly on a daily basis.

Geoff allowed us sufficient flexibility in focussing on areas of particular interest of our own.  I began my work when Geoff suggested that I take a look at various questions of bounds, and, in particular, the question of the Cerulus--Martin lower bound~\cite{Tan:1967a}, which had just gained certain interest from an experimental perspective. Asymptotic behavior of scattering amplitudes has already been recognized as an integral part of S-matrix consistency.  The Regge hypothesis, which had become popular due to its application to experiment,  was  introduced as an integral component of bootstrap consistency\cite{Chew:1963zza}.  

I was fortunate that Prof.~Richard Eden of Cambridge came to Berkeley for a year's sabbatical, and I worked with him closely, which formed the main part of my thesis. Although my thesis work  seemed formal, it connected well with the common interest of  all my fellow students, as well as a large group of post-docs.  Although we each had our own research focus, we all shared the excitement of the bootstrap philosophy, with regular weekly informal seminars.   

The group grew steadily in size and one often could not find seats in the Building-50 seminar/coffee room.
When Geoff and Ruth moved back to the Berkeley hills, Geoff invited his group of students for an evening get-together at their beautiful new home. As a house-warming gift, I organized and brought Chinese take-outs.  It was a wonderful evening, the discussion invariably turning to physics, while we enjoyed the view of the Campanile and Golden Gate in the distance. 
 Whether by design or not, this informal gathering turned into a regular weekly meeting in the Berkeley hills for his students.\footnote{In addition to  discussing physics, other stories were also exchanged, e.g., Geoff's recollections of his student days at Chicago with Fermi, Goldberger, Yang, Lee, etc. I also remember his story about meeting Landau, who reminded Geoff of the significance of his being the first to emphasize the importance of the particle--pole correspondence.} I understand this practice continued after those first timers had left Berkeley after receiving their PhDs.

In looking back one cannot avoid concluding that the Pomeron remained the central topic of interest to most of us at that time.  In this remembrance of Geoff, I will discuss how the issue of the Pomeron evolved over the past 50 years.

\section{Rencontres de Moriond and the Soft Pomeron}

The ``Rencontres de Moriond"  is a series of annual high energy physics meetings starting in 1966. The initial impetus for this series is ``to promote fruitful collaboration by bringing together a small number of scientists in inspiring surroundings, to discuss recent findings and new ideas in physics."
This series was  organized by Prof.~Tran Thanh Van of the University of Paris, Orsay.  It  has continued to thrive over more than half a century. As I will explain below for the period of 70s and 80s, Pomeron Physics  suffered a period of ``crisis." However, during this period, these annual meetings  served as a playground where the physics of the Pomeron flourished, from the experimental perspective. In particular, it provided support for the notion that the Pomeron is non-perturbative and exhibits aspects of string structure with a cylindrical topology, i.e., the exchange of a closed string.

\vspace{-2mm}

\paragraph{Crisis for the Pomeron:}

There are several developments in the late 60s and early 70s which dealt unkindly to the study of the Pomeron. (a) Since total cross sections can be interpreted geometrically, where does the scale come from for the Pomeron coupling?
 (b)  Total cross sections continued to increase with energy, albeit slowly. This is in conflict with the initial impetus of having the Pomeron with an exact intercept at $j=1$. (c) As a collection of closed string excitations\cite{GV68,SS}, flat-space string theories led to a Pomeron intercept at $j=2$.  What are the low-lying particles  on this trajectory? (d) Lastly, what is the Pomeron  in  QCD?  

Let me comment on these aspects briefly, in reverse. QCD is a theory of quarks and gluons. Although confinement in the infrared (IR) remains to be demonstrated explicitly, most theorists believe this can eventually be accomplished. This belief is further supported by lattice studies, where the low-lying particle spectrum has been calculated.  Presumably, particles on the Pomeron trajectory should be identified with a tensor glueball and its recurrences.  As a non-perturbative phenomenon, the Pomeron is not considered  a subject of immediate concern. 

Having an intercept $j=2$ for closed string excitations has led to the possibility of a theory for quantum gravity.  This, of course, is based on string theory in flat space--time, albeit at a higher dimension. More seriously,  string theories also suffer difficulty in introducing local currents. As a consequence, it is difficult to see how the experimental observation of power-behaved amplitudes at large momentum transfer can be accommodated.

\paragraph{The Topological Pomeron and Bootstrap:} 
It has become increasingly clear that the increase in the total cross sections is driven by an increasing rate of inelastic particle production. Since particle production can be successfully explained by a Regge mechanism, it remains consistent to assume that the Pomeron, through unitarity, should emerge as a sum of ladder-like diagram, e.g., so-called multiperipheral mechanism~\cite{Chew:1969nn}. Indeed, this picture is consistent with string theories, with the Pomeron entering at a higher genus, e.g., exchanging a Pomeron has the topology of a cylinder.  This picture was advocated by Veneziano, leading to the program of ``dual topological unitarization"\cite{Veneziano:1971fp} (DTU). (See also the contribution by Veneziano in this volume.)

From a QCD perspective, one optimistically  awaits for a future QCD string theory, propagating over a curved background. In the meantime, it is important to  continue to search for evidence for such a topological structure via particle production. Notably, in a topological expansion, the Froissart bound no  longer applies at the cylinder-level, which removes the proposition that   the Pomeron has a unit intercept. In fact once this restriction is removed, the value of the Pomeron intercept, being greater than unity, becomes a challenging dynamical question.  Experimentally it seems to be greater by an amount of the order of $0.1\sim 0.2$.  Can one provide an upper bound? Finally, the Froissart bound should ultimately be restored, e.g., via eikonalization. At what level should a geometrical scale enter in the discussion?

\vspace{-2mm}

\paragraph{DTU:}
Much of particle production data during this period came from experiments at CERN. It was natural that new data was presented and discussed at Moriond, which was then held in the French alps.  Due to its close proximity to Geneva, many theorists at CERN would regularly participate in these meetings, e.g., G.~Veneziano, A.~White, D.~Amati, Chan Hong-Mo, etc. Theoretical  interpretation for particle production was a regular topic of intense discussion.

\begin{figure}[ht]
\centering
\includegraphics[width=5 in]{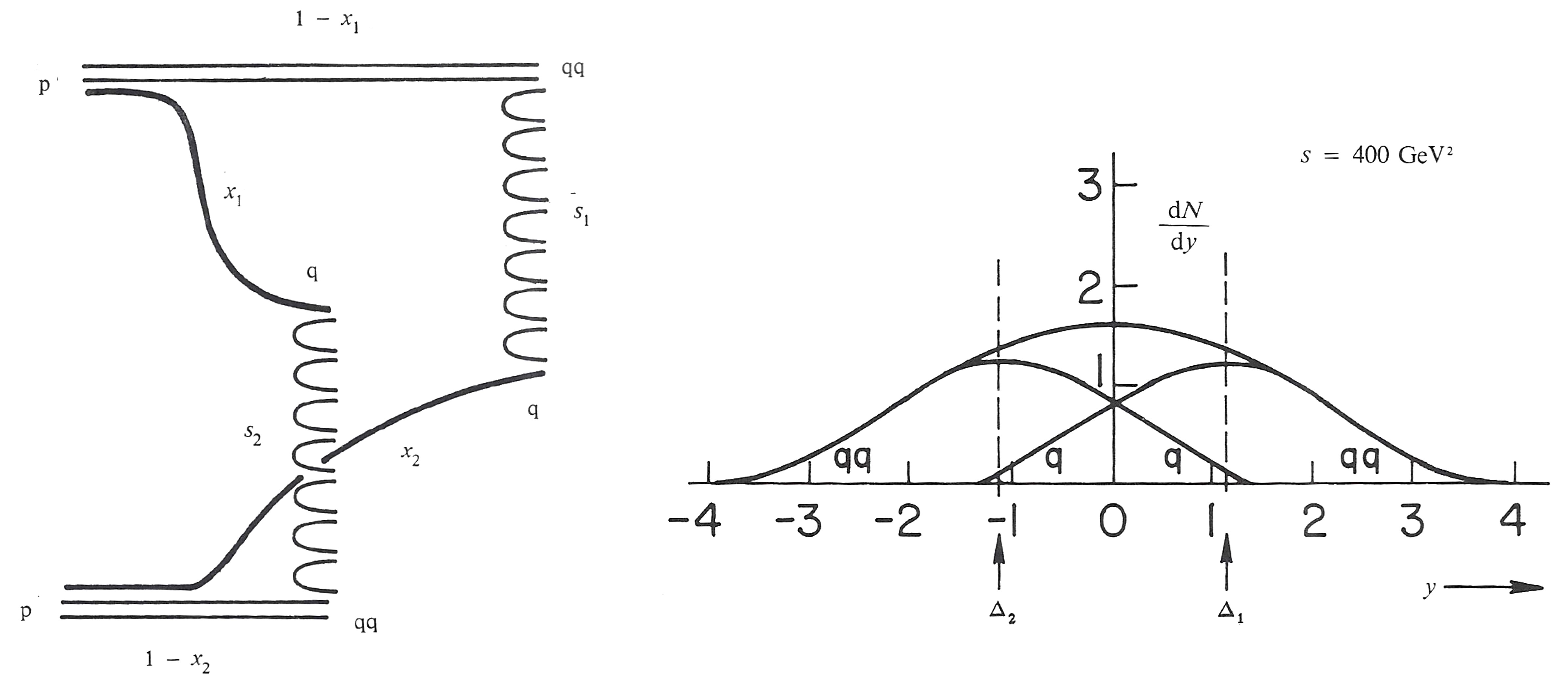}
\caption{A cut-topological Pomeron for particle production and the resulting inclusive cross section.}
\label{fig:delta-j}
\end{figure}

During my first sabbatical at Orsay, working with A.~Capella, U.~Sukhatme, and Tran Thanh Van, we began an exploration of the consequences of the Pomeron having the desired topological structure, as suggested by string theory in general and DTU in particular. These studies were vigorously discussed and confronted with experiments. In particular, these were carried out regularly at annual Moriond meetings. Based on the initial success, a systematic calculus was formulated, which has subsequently dubbed the ``Dual Parton Model" (DPM)\cite{Capella:1992yb}. This is partly based on the general approach to the S-matrix bootstrap advocated by Veneziano~\cite{Veneziano:1971fp} and also adopted by Geoff~\cite{Chew:1977yk}. The details of this model were discussed in the contribution {\it The Pomeron Story}, and  will not be repeated here. (See Ref.~\cite{Capella:1992yb}  and also the contribution of \mbox{G.~Veneziano} in this volume.) Here we illustrate the key features by two schematics  in Fig.~\ref{fig:delta-j}.

\vspace{-1mm}

\noindent Through these efforts, the notion of a topological Pomeron was firmly established phenomenologically. Since this identification is established via the production of particles, which tend to have limited transverse momenta, it has been dubbed a \mbox{``soft Pomeron."} This usage also indicates that further complications are lurking around the corner and would soon appear in the coming years.

\section{A Perturbative  vs a Non-Perturbative Pomeron}

As  hinted above, it soon became  clear that jet production would  become an important component of particle production. The question can then be posed, how to  account for the total cross section resulting from such processes. However, it is unclear if this separation can be made meaningfully before a fully comprehensive confinement picture emerges. In any event, both during Moriond and also at other meetings,  this issue became increasingly the center of discussion. One began to talk about ``hard-Pomeron" versus ``soft-Pomeron." Loosely speaking, hard-Pomeron should be calculable based on a perturbative approach, while soft-Pomeron remains non-perturbative. Should the bootstrap be restricted to the soft component only?

\paragraph{BFKL:}
Beginning in the mid-1970s, Lev Lipatov~\cite{Lipatov} and his collaborators (the Leningrad school),  carried out  a series of studies of the high-energy behavior of scattering amplitudes in QCD. These were serious and difficult analyses, which ultimately led to the emergence of a Pomeron-like structure. However, this study was carried out mostly by assuming conformal invariance, leading to a Pomeron being a branch-point, instead of a pole. Nevertheless, the location of this singularity was calculated. To first order in 
the 't Hooft  coupling, $\lambda$, it is located above unity, 
$$
\alpha_P= 1 + (\ln 2/\pi^2) \lambda \,.
$$
It should be emphasized that this result follows from summing gluon ladder graphs, which clearly ignores confinement: thus it is perturbative.  Nevertheless, it has resulted in a robust research program, including an application to particle production. Furthermore, the BFKL program was able to lead to interesting predictions for deep-inelastic scattering (DIS) experiments at HERA.  
 However, this picture clearly runs counter to that for DTU, responsible for a {\it soft Pomeron.} The corresponding BFKL Pomeron, is now identified with  the {\it hard Pomeron}.

\paragraph{Donnachie--Landshoff:}
While the work of BFKL received  increasing   attention from  both theoretical and experimental communities, a surprisingly simple parametrization was proposed by Donnachie and Landshoff for a soft-Pomeron~\cite{D-L}.  They were able to explain experimental data for total cross sections for $\bar pp$, $pp$, $\pi^-p$, $\pi^+p$, $K^-p$, $K^-p$, and $\gamma p$ over a large energy range. It was a truly impressive fit to the data, with an effective Pomeron intercept of
$$
\alpha_P\simeq 1.08 \,.
$$
This value is significantly lower than that expected for the BFKL-Pomeron. This analysis is purely phenomenological, (For more discussion of this development, see the contribution by Donnachie and Landshoff in this volume.)   No suggestion  on how this trajectory can be generated dynamically was provided,  e.g., what states lie  on this trajectory?

\paragraph{Russians are Coming --- Moriond 1992:}
While the late 1960s was the period of social unrest and change in the US, the same can be said for the Soviet Union in the 1980s.  Moriond had always been welcoming to physicists  from eastern block  countries. Beginning in late 1980s, increasingly more participants were able to come from the former Soviet Union. In particular, many from the Leningrad school began to come regularly. In 1992, while again on sabbatical at Orsay, I was fortunate to meet Genya Levin at Moriond. 
Genya is a contemporary of Lev Lipatov; both were students together working with Gribov. While I tried to convince him of the success of DTU, leading to a string-like topological-Pomeron, Genya would argue forcefully that the correct approach should be BFKL-Pomeron~\cite{Levin:1976mc}. 

After several days of argument, we both began to see  the values of the opposite view.  We clarified for each other how each respective Pomeron can be understood in terms of the picture of random walks. The difference between the two lies in the respective space in which this takes place. In the case of the Soft-Pomeron, the random walk occurs in the transverse impact space. In contrast, the BFKL Pomeron is generated primarily by random walks in ``virtuality."

We kept up our communication after Moriond, and soon became good friends. We also came to the conclusion that a proper  treatment of high energy scattering should incorporate both soft- and hard-features. We soon came up with a toy-model which would do precisely that. We coined the phrase ``Heterotic Pomeron," and I presented a talk on this proposal at the July 1992 meeting for International Symposium on Multiparticle Dynamics (ISMD)  at Santiago, Spain, with the intention of writing up a formal paper shortly afterwards.  For various reasons, the project was delayed. (We finally were able to finish this project nearly ten years later~\cite{Levin:1992ys}.)

\section{The Pomeron and Gauge--String Duality}

A major development in our understanding of non-pertubative gauge theory occurred in 1997, the AdS/CFT conjecture of Maldacena.  In particular, motivated by a suggestion of Witten, Jevicki {\it et al.} and independently  Csaki {\it et al.} showed how scalar glueball masses can be calculated via AdS/CFT, once  confinement is implemented.  Working with  Richard Brower, together with a then young colleague at MIT, Samir Mathur, we were able to carry out an analogous analysis generalized to tensor glueballs~\cite{Brower:2000rp}. This is by itself an interesting exercise, which I learned greatly from Rich and Samir. Equally important is a paper by Polchinski and Strassler where they explained how point-like power behavior for scattering at large angle can emerge when AdS/CFT is implemented.  
Secretly, I was extremely excited; at long last, a promising theoretical framework lay ahead as the basis for the phenomenologically motivated heterotic Pomeron. 

In 2007, Brower and I,  together with J.~Polchinski and M.~Strassler\cite{Brower:2006ea},
 provided  a coherent first principles treatment of the Pomeron. In the large-N QCD-like theories, we used curved-space string-theory to describe simultaneously both the BFKL regime and the classical Regge regime. On  the basis of gauge/string duality,  this allows us to deal with  high-energy small-angle scattering in QCD.  The problem reduces to finding the spectrum of a single j-plane Schr\"odinger operator.  Our results agree with expectations for the BFKL Pomeron at negative t, and  with the expected glueball spectrum at positive t, but provide a framework in which they are unified.  
 
 In short, a dual-topological Pomeron is nothing but a reggeized graviton over a curved background, with the Pomeron intercept depending on the curvature.  To first order in $1/\sqrt \lambda$,  we find
$$
\alpha_P\simeq 2-2/\sqrt \lambda \,.
$$
This gauge-string motivated Pomeron has been referred to by others as the ``BPST Pomeron''. It is  heterotic,  as previously proposed. It is based on  firm principles, and should  be contrasted with weak-coupling BFKL Pomeron. 

Further applications of this dual-topological  Pomeron has been carried out in recent years~\cite{Brower:2010wf}. Effects beyond the single Pomeron exchange have also been  discussed\cite{Brower:2007xg}. Future developments and challenges undoubtedly await, e.g., the recent reported observation of the Odderon~\cite{Martynov:2018yas}  requires further clarification from a string-dual perspective~\cite{Brower:2014wha}. Similarly, the influence of the pion mass on the total and elastic cross sections has also been raised in the context of the topological Pomeron~\cite{Tan:1710.10594}. More intriguingly, can the value of the Pomeron intercept be fixed by the bootstrap? 

\section{Moriond 1984}

After Berkeley, I  moved to Princeton in 1968 as  a post-doc, working primarily with Murph Goldberger, in particular on understanding the Pomeron from the perspective of the multiperipheral picture. After completing my post-doc, I moved to Brown as an Assistant Professor in 1970 and stayed since. Geoff and I kept up with our interactions during this period --- in particular in  1970 when Geoff spent his sabbatical at Princeton.   Equally rewarding  is the fact that I was able to work with Carleton DeTar and Joe Weis~\cite{Detar:1971dj}, my fellow graduate student colleagues at Berkeley, who had just moved to MIT. We had a fruitful collaboration, making use of Veneziano models for inclusive production via the multi-Regge picture. This is an extension of the work of Mueller, which strengthened the usefulness of a string-like picture for particle production.

I spent a fruitful   sabbatical  at Orsay in 1976--77, working with Capella, Sukhatme and Tran on DTU.  After my return to Brown, my interests gradually shifted to the study of  QCD non-perturbatively via the large-N expansion. I was greatly helped by my colleagues at Brown, in particular A.~Jevicki, who introduced me to the collective-field approach. I was also fortunate in the following year to team up with  my long-time graduate student friend, Rich Brower, who was spending a year at Harvard,  in the study of matrix models.  Together with a young post-doc at MIT, P.~Rossi,  we worked out and extended several nontrivial examples of coupled matrix models~\cite{Brower:1980vm}. Rich also introduced me to the detailed working of lattice gauge theory, for which I am truly grateful, and he helped direct my graduate student, Kostas Orginos, who is by now an accomplished  lattice gauge theorist at William and Mary.  During this period, I did not have much direct contact with Geoff, although I have tried to keep up with his work when preprints arrived.

I  had the fortune to overlap with Geoff again in 1984 when I spent my second sabbatical at Orsay.  However, Geoff and I were located at different research institutes, and we ate at different canteens, preventing us from having regular get-togethers. At the same time my research interests  had also diverged from his. I began to engage increasingly more with model studies in QCD at large N, while maintaining my continuing interests in phenomenological applications to multiparticle production based on the topological Pomeron. At the same time, Geoff focused on developing a topological approach to the S-matrix bootstrap~\cite{Chew:1981hh}. 

While he would occasionally listen to our explanations about how the Pomeron would fit in a topological setting, clearly he was less and less interested in phenomenological applications. Although we did not meet on a daily basis at Orsay, we tended to often  catch the same commuter train back to Paris. During the short 30-minute ride, he would, in his usual enthusiastic fashion, expound on his latest thinking on physics. While I don't have a specific recollection of these conversations, I do remember one incident of interest: During one of the return trips, while we continued to chat at the transfer point at Denfert-Rochereau, Geoff realized that his commuter ticket was missing. After I had passed through the gate, Geoff, while looking briefly left and right for security guards, simply jumped over the barrier with a quick athletic move.  We each moved quickly to the next quay to catch our respective transfers, Geoff with a smug smile on his face.

My last extended interaction with Geoff took place at the 1984 Moriond conference at les Arcs 1600, France. I recall Geoff expounding on his optimistic view of how unification of the standard model would emerge from a topological bootstrap approach~\cite{Chew:1981hh}. He was at his usual best, trying to share his excitement on how new physics can be examined from a bootstrap  perspective. After the 1984 Moriond meeting, my interactions with Geoff came less often. 

\vspace{1mm}

\section{Last  Visits}
 
I visited  Geoff in 2010 while attending a conference in Berkeley. I had lunch with Geoff at the usual Rad Lab cafeteria, together with Geoff's longtime collaborators, Henry Stapp and Jerry Finkelstein.   After lunch, we returned to their office, and Geoff tried to explain to me their latest research efforts. Although the visit was short it brought back memories of the wonderful graduate student days.

I had another chance to 
visit Geoff  in September 2013. We had lunch together at his home in the Berkeley hills. How it brought back memory of fifty years ago when we went there for the first time after the house was built! Instead of Chinese take-out Geoff had ordered pizzas for lunch. He was pleased when I shared with him the resurgence of the Bootstrap program in current CFT studies as well as the heavy reliance on S-matrix principles for amplitude studies. I also updated him on the status of our topological Pomeron program. I learned from Geoff about his current goal in understanding cosmology from a bootstrap perspective~\cite{Chew:2013qfa}. He also talked more about his graduate student years with Fermi. I was also  fascinated later by the  story of  how he had interacted with Von Neumann as an undergraduate at George Washington University and how G.~Gamov likely helped start his physics career by sending him to join the Manhattan Project. When it was time to leave I had no idea that would be the last time I would see him.

\begin{figure}[h]
\centering
\includegraphics[width=4.4 in]{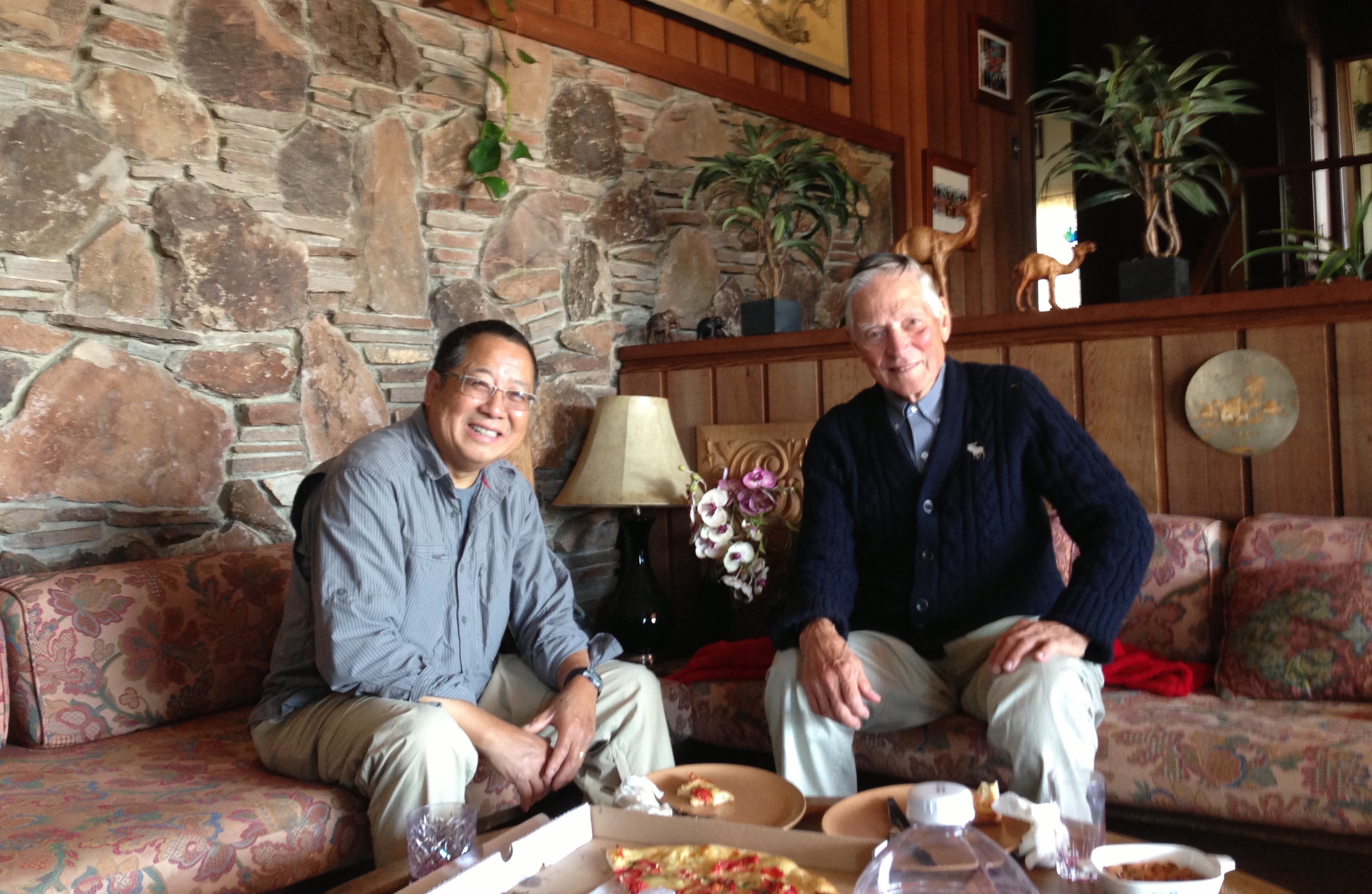}
\caption{Photo taken during my 2013 visit.}
\label{fig:chew-tan}
\end{figure}

\vspace{-2mm}

In March 2018, 
we had a chance to visit DAMTP where my former student, Kostas Orginos, was on sabbatical there. His wife, Lily, a classicist, was also on sabbatical at Cambridge.  Through her connection they were staying at Clare Hall, where Richard Eden  is a founding Fellow.   The highlight of our visit was to have lunch with Richard. It was a pleasure to listen to him re-telling the story of his graduate days working with Dirac.

 \fontdimen2\font=3.25pt

I started my research career more than 50 years ago  by trying to understand the Pomeron and the Bootstrap.  It remains a personal challenge and {\it The Pomeron Story} continues.
 It has been a wonderful and continuing journey. Thanks to Geoff, my fellow graduate students and Eden~\footnote{ Richard Eden passed away in 2021, shortly before the completion of this book.}, for  shaping my research interest and trajectory.   

 \fontdimen2\font=3.5pt

\end{document}